# Alias and Change Calculi, Applied to Frame Inference

Alexander Kogtenkov[1, 3], Bertrand Meyer[1, 2, 3], Sergey Velder[3]

[1] Eiffel Software (Santa Barbara)  [2] ETH Zurich  [3] NRU ITMO (Saint Petersburg)

alexk@eiffel.com, velder@rain.ifmo.ru, Bertrand.Meyer@inf.ethz.ch

**Abstract.** Alias analysis, which determines whether two expressions in a program may reference to the same object, has many potential applications in program construction and verification. We have developed a theory for alias analysis, the "alias calculus", implemented its application to an object-oriented language, and integrated the result into a modern IDE. The calculus has a higher level of precision than many existing alias analysis techniques.

One of the principal applications is to allow automatic *change analysis*, which leads to inferring "modifies clauses", providing a significant advance towards addressing the Frame Problem. Experiments were able to infer the "modifies" clauses of an existing formally specified library. Other applications, in particular to concurrent programming, also appear possible.

The article presents the calculus, the application to frame analysis including experimental results, and other projected applications. The ongoing work includes building more efficient model capturing aliasing properties and soundness proof for its essential elements.

## 1 Overview

A largely open problem in program analysis is to obtain a practical mechanism to detect whether the run-time values of two expressions can become *aliased*: point to the same object. "Practical" means that the analysis should be:

- *Sound*: if two expressions can become aliased in some execution, it will report it.
- *Precise* enough: since aliasing is undecidable, we cannot expect completeness; we may expect false positives, telling us that expressions may be aliased even though that will not happen in practice; but there should be as few of them as possible.
- *Realistic*: the mechanism should cover a full modern language.
- *Efficient*: reasonable in its time and space costs.
- *Integrated*: usable as part of an integrated development environment (IDE), with an API (abstract program interface) making it accessible to any tool (compiler, prover…) that can take advantage of alias analysis.

The present discussion considers "may-alias" analysis, which reports a result whenever expressions *may* become aliased in *some* executions. The "must-alias" variant follows a dual set of laws, not considered further in the present paper.

The papers [11, 12] introduced the **alias calculus**, a theory for reasoning about aliasing through the notion of "alias relation" and rules determining the effect of every kind of instruction on the current alias relation. We have refined, corrected and extended the theory and produced a new implementation fully integrated in the EVE (Eiffel Verification Environment) open-source IDE [1] and available for download at the given URL. In the classification of [6, 7, 18], the analysis is untyped, flow-sensitive, path-insensitive, field-sensitive, interprocedural, and context-sensitive.

The present paper describes the current state of alias analysis as implemented. It includes major advances over [12]:

- The calculus and implementation cover most of a modern OO language.
- The implementation is integrated with the IDE and available to other tools.
- The performance has been considerably improved.
- A large part of the calculus has been proved sound, mechanically, using Coq.
- An important error affecting assignment in an OO context that has been corrected.
- New applications have been developed, in particular to *frame analysis*.

Frame analysis relies on a complement to the alias calculus: the *change calculus*, also implemented, which makes it possible to infer the "modifies clause" of a routine (the list of expressions it may modify) automatically. Applied to an existing formally specified library including "modifies" clauses, the automatic analysis yielded all the clauses specified, and uncovered a number of clauses that had been missed, even though the library, intended to validate new specification techniques (theory-based specification), has been very carefully specified.

Section 2 presents the general assumptions and section 3 the calculus. Section 4 introduces the change calculus and automatic inference of frame conditions. Section 5 describes the implementation and the results it yielded in inferring frame conditions for a formally specified library. Section 6 presents the ongoing work concerning other applications, such as deadlock detection, and a new theoretical basis. Section 7 discusses related work. Section 8 is a conclusion and review of open problems.

## 2    The mathematical basis: alias relations

$E$ denotes the set of possible expressions. An expression is a path of the form *x.y.z…* where *x* is a local variable or attribute of one of the classes of the program, or *Current*, and *y*, *z*, …, if present, are attributes. Variables and attributes are also called "tags". *Current* represents the current object (also known as "this" or "self".)

An alias relation is a binary relation on $E$ (that is, a member of $\mathbb{P}\,(E \times E)$) that is symmetric and irreflexive. If $r$ is an alias relation and $e$ an expression, $r\,/\,e$ denotes the set made of all elements aliased to $e$, plus $e$ itself: $\{e\} \cup \{x\colon E \mid [x, e] \in r\}$. An alias relation may be infinite; for example the instruction $a.set\_u\,(a)$, where $set\_u$ assigns the $u$ field, causes $a$ to become aliased to $a.u.u….$, with any number of occurrences of the tag $u$; in this case the set $r\,/\,x$ of aliases of $x$ will be infinite.

Alias relations are in general not transitive, since expressions can receive different aliases on different branches of a program: **if** $c$ **then** $x := y$ **else** $x := z$ **end** yields an alias relation that contains the pairs $[x, y]$ and $[x, z]$ but not necessarily $[y, z]$.

To define the meaning of alias relations, we note that the calculus cannot be complete, since aliasing is undecidable for a realistic language. It must of course be sound; so the semantics (section 6.3) is that if an alias relation $r$ holds in a computation state, then any pair of expressions $[e, f]$ *not* in $r$ is *not* aliased (i.e. $e \neq f$) in that state.

A convenient way to write an alias relation is the **canonical form** $A$, $B$, $C$, … where each element is a set of expressions $e$, $f$, …, none of them a subset of another; such a set is written $\overline{e, f, \ldots}$. For example the above conditional instruction, starting from an empty alias relation, yields $\overline{x, y}$, $\overline{x, z}$. More generally $\overline{A}$, for a list or set of expressions $A$, denotes $A$ symmetrized (by adding $[y, x]$ whenever $[x, y]$ is present) and "de-reflexived" (by removing any pair $[x, x]$).

## 3  The alias calculus

The alias calculus is a set of rules defining the effect of executing an instruction on the aliasings that may exist between expressions. Each of these rules gives, for an instruction $p$ of a given kind and an alias relation $r$ that holds in the initial state, the value of $r \gg p$, the alias relation that holds after the execution of $p$.

By itself the alias calculus is automatic: it does not require programmer annotations. Since it only addresses a specific aspect of program correctness, it may have to be used together with another technique of program verification, in particular Hoare-style semantics, which uses annotations. The relation goes both ways:

- If a routine's postcondition expresses a non-aliasing property $x \neq y$, the calculus can prove it (using lighter techniques than the usual axiomatic proof mechanisms).
- Conversely, the alias calculus may need to rely on properties established separately. In particular, it ignores conditional expressions; so in computing $r \gg$ (**if** $x \neq y$ **then** $z := x$ **end**) where $r$ contains $[x, y]$, it will yield a relation containing $[x, z]$ even though $x$ and $z$ cannot actually become aliased. In many cases the resulting imprecision is harmless, but its removing requires help from other techniques. The solution takes the form of an instruction **cut**, such that $r \gg$ (**cut** $x, y$) is $r$ deprived of the pair $[x, y]$ and all pairs $[x.e, y.e]$.

The **cut** instruction is an annotation added by programmers when they need more precision. In the interplay between the alias calculus and a Hoare-style prover, **cut** $x, y$ is a benefit for the calculus and a proof obligation for the prover. We have not encountered the need for such annotations so far, but they provide a safety valve.

Since the calculus ignores tests, we omit them in conditionals, writing them just **then** $p$ **else** $q$ **end**, and loops, written **loop** $p$ **end**. The rules for control structures are:

$r \gg (p \, ; q) = (r \gg p) \gg q$

$r \gg$ (**then** $p$ **else** $q$ **end**) $= r \gg p \cup r \gg q$

$r \gg$ (**loop** $p$ **end**) $= t_N$, for the first $N$ such that $t_N = t_{N+1}$,

— where $t_0 = r$ and $t_{n+1} = t_n \cup (t_n \gg p)$ (see below about finiteness)

For a creation instruction ($x := new$ (…) in Java style) and a "forget" ($x := null$):

$r$ » (**create** $x$) = $r - x$
$r$ » (**forget** $x$) = $r - x$

where "$-$" is set difference generalized to elements ($x$ stands for $\{x\}$), relations and paths: $r - x$ is $r$ deprived of all pairs of which one element is $x.e$ (or $e_0.x.e$ where $e_0$ is aliased to *Current* in $r$). For "**cut**" we have:

$r$ » (**cut** $x$, $y$) = $r - \overline{x, y}$

The rule for unqualified routine call, with $l$ as the list of actual arguments, is:

$r$ » (**call** $f(l)$) = ($r$ [$f^\bullet$: $l$]) » $|f|$

where $f^\bullet$ is the formal argument list of $f$, $r$ [$u$: $v$] the relation $r$ with every element of the list $v$ substituted for its counterpart in $u$, and $|f|$ the body of $f$.

The rule for qualified calls relies on a notion of "negative variable" [12, 13] to transpose the context of the call to the context of the caller:

$r$ » ($x$ **.** **call** $f(l)$) = $x$ **.** (($x'.r$) » **call** $f(x'.l)$)

where $x'$ is the "negation" of $x$, with $x'.x = Current$, and "**.**" is generalized distributively to lists ($x.\langle a, b, …\rangle = \langle x.a, x.b, …\rangle$) and relations.

The main instruction that creates aliasings, removing previous ones, is reference assignment: $t := s$. The assignment rule given in [12] was unsound (in the cases when any expression of $r / s$ starts from expression of the form $e.t$ where $e$ is aliased to *Current* in $r$). The new rule has been proved sound in semantics discussed in section 6.3. It can be expressed in several ways, of which the easiest to understand uses a fresh variable $ot$ (for "**old** $t$"):

$r$ » ($t := s$) = **given** $r_1 = r$ [$ot = t$] **then** ($r_1 - t$) [$t = (r_1 / s) - t$] $- ot$ **end**

with $r$ [$x = u$] denotes the relation $r$ augmented with pairs [$x, y$] where $y$ is an element of $u$, and made *dot complete* (in terms of [12]), that is to say, for any $t$, $u$, $v$ and $a$ if [$t, u$] and [$t.a, v$] are alias pairs, then also [$u.a, v$], and if [$t, u$] is alias pair and $a$ is in the domain of $t$, then [$t.a, u.a$] is alias pair. The set difference "$-$" is defined above.

Since alias analysis cannot be complete, the calculus introduces possible imprecisions (over-approximations); it is important to understand where they actually lie. In fact, the above rules are precise. Over-approximations come from ignoring conditions in conditionals and loops, such as $c$ in **if** $c$ **then** $a$ **else** $b$ **end**. It is possible to remove any such imprecision by introducing **cut** instructions, but these have to be established through separate means, such as a Hoare-style prover, outside of the alias calculus.

The implementation of the calculus introduces another source of possible imprecision. In an OO language with unbounded run-time object structures, the alias relation may, as noted, be infinite. To stick to finite structures the implementation must cut off the graph. The first step [12] is to limit ourselves to $M$, the maximum length of a path appearing in an expression of the program (including contracts, especially postconditions). This is, however, not sufficient; in a case such as:

$a := first$;   $a := a.right$; $a := a.right$; … — $n$ times
$b := first$;   $b := b.right$; $b := b.right$; … — $n$ times

where $n > M > 1$, the expressions $a$ and $b$, both of length $< M$, become aliased to each other through being both aliased to an expression of length greater than $M$ that does not appear in the program: *first.right.right…* (*n* "*right*" tags). A similar problem arises for code containing loops:

*a* := *first*; **loop** *a* := *a.right* **end**;
*b* := *first*; **loop** *b* := *b.right* **end**;

The implementation and the formal model use a maximum path length $L \geq M$ and treat any expressions longer than $L$ as aliased to all expressions. This technique introduces imprecision but retains soundness. In the future it may be improved using type information (in a statically typed language *e* and *f* can only be aliased if their types are compatible; also in polymorphic version of the qualified call rule we replace the resulting alias relation by the union of similar alias relations for all features corresponding to inherited classes). Unlike some of the approximations found in the alias analysis literature, where the equivalent of $L$ is very small, our $L$ can run into large values.

## 4    The change calculus and frame condition inference

One of the key problems of software verification, still largely open for OO programs, is frame analysis: determining what an operation does *not* change. Current solutions, following in part from tools such as ESC/Java [2] and its successors, assume that the programmer writes a "modifies clause" listing the expressions whose value may change. (As a matter of syntactic taste we prefer the keyword "**only**" to "**modifies**", since the goal is not to list expressions that *will* change, but to specify that any expression not listed will not change.) Writing such clauses is, however, tedious. It is hard enough to convince programmers to state what their program does; forcing them in addition to specify all that it does not do may be a tough sell. It is desirable, as much as possible to infer the "modifies" clauses.

The alias calculus opens the way to such an approach by enabling a *change calculus* (an abbreviation for *may-change* calculus) which, for any instruction *p*, yields $\underline{p}$, the set of expressions whose value may change as a result of executing *p*. Like the alias calculus, the change calculus is an over-approximation: for soundness $\underline{p}$ must include anything that changes, but conversely an expression might appear in $\underline{p}$ and not change in some executions of *p*, or even (as a sign of our incompetence, inevitable because of undecidability) in none of them. The basic rules of the calculus are (*r* is the alias relation in the initial state, $r / x$ is the set of aliases of *x* plus *x* itself, and "**.**" distributes over sets):

| | | |
|---|---|---|
| $\underline{t := s}$ | $= (r \,/\, Current) \,.\, t$ | |
| $\underline{p \,;\, q}$ | $= \underline{p} \cup \underline{q}$ | |
| **then** $\underline{p}$ **else** $\underline{q}$ **end** | $= \underline{p} \cup \underline{q}$ | — same as for ";" |
| **loop** $\underline{p}$ **end** | $= \underline{p} \cup \underline{p}^2 \cup \underline{p}^3 \cup \ldots$ | — limited to $L$ elements as discussed |
| **call** $\underline{f\,(l)}$ | $= \lfloor f \rfloor \,[l \colon f^\bullet]$ | |

The most important rule, requiring alias analysis, is for qualified calls:

**call** $\underline{x.f\,(l)} = (r \,/\, x) \,.\, (\textbf{call}\ \underline{f\,(x'.l)})$

where, as before, "**.**" distributes over lists and relations and $y.x' = Current$ if *x* and *y* are aliased in *r*. The rule states that for any *u* that *f* may change, **call** $x.f\,(l)$ may change not only *x.u* but also *y.u* for *y* aliased to *x*.

The change calculus, implemented on top of the alias calculus thanks to this rule, enables us to infer frame conditions. This inference is a possible over-approximation. It makes it possible to verify programmer-supplied "modifies" clause in the following way. Let $p_c$ be the set of expressions that can change as a result of the execution of an instruction $p$, typically a routine call. Let $p_m$ be the list of expressions in the "modifies" clause. The clause is sound if and only if

$$p_c \subseteq p_m \qquad \text{[C1]}$$

For theoretical reasons (undecidability) and practical ones (tool limitations), the verification cannot compute $p_c$ exactly; instead it computes $\underline{p}$. Assuming soundness of the change calculus (and hence of the alias calculus), we have the guarantee that

$$p_c \subseteq \underline{p} \qquad \text{[C2]}$$

In other words, $\underline{p}$ is a possible over-approximation of the actual change set. Then if a tool such as our implementation is able to compute $\underline{p}$, a compiler can examine the program and its annotations to ascertain the property

$$\underline{p} \subseteq p_m \qquad \text{[C3]}$$

which guarantees [C1] and hence the correctness of the "modifies" clause.

In our work towards an integrated development and verification environment, EVE, we intend, for the reasons mentioned above, not to include syntactic support for a "modifies" (or **only**) clause. Instead we simply consider that any expression not listed in the postcondition (**ensure**) of a routine is considered to be unchanged by the routine. This approach is validated by an informal survey of specifications in JML libraries, which suggests that in the practice of specification every expression $e$ listed in a "modifies" clause *also* appears in the postcondition. For any exceptions to this observation it is always possible to include a special predicate *involved* (*e*).

This convention has not yet been applied on a large scale. Until it is, we are validating the calculus on code with explicit "modifies" clause, as discussed in section 5.

## 5 Implementation, and results of frame analysis

The alias and change calculi described in previous sections have been fully implemented. Earlier papers [11, 12] described a prototype stand-alone implementation. The present implementation is integrated in EVE [1], the research version of EiffelStudio, a modern integrated IDE covering the full Eiffel language. The execution time of the analysis in our experiments is (on an ordinary laptop) about 7 minutes for a mid-sized library class. We are working to improve the performance so as to allow immediate feedback; it is, however, acceptable for the current uses of the analysis in the EVE multi-tool, multi-method verification environment, where cooperating verification tools running in the background deliver diagnostics and suggestions to the user.

To assess the approach we performed change analysis on a formally specified library, EiffelBase+ [16]. The library has the attraction of providing "*full contracts*" that specify all properties; for example the postcondition of a "push" operation for stacks states not only that the number of elements has been incremented by one and that the new top is the routine's argument, but also that the previous elements remain.

EiffelBase+ currently includes "modifies" clauses. Since the specification style relies on mathematical "model queries" (theory-based specification), these clauses list such queries, not directly the attributes (fields). An example model query, for class *STACK*, is *sequence*, which gives the associated sequence of elements. Running the analysis required mapping attributes to model queries. In most cases the correspondence is straightforward: many model queries map directly to attributes. In a few cases, the model query has no direct attribute counterparts; for example, the model query *sequence* of *LINKED_LIST* is computed by traversing all elements of the list.

We ran the frame inference on 36 classes with 278 "modifies" clauses, detecting a number of missing or different "modifies" specifications; for example, the analysis reports that routines *disjoint* and *is_subset* of a class *ARRAYED_SET* can modify the attribute *index*, not listed in the "modifies" clause. The full results with detailed analysis of found differences are available at sel.ifmo.ru/results/alias/EiffelBase+/.

For 614 analyzed features, 592 (96%) "modifies" clauses could be mapped from model to source code. For that code the analysis yielded 100% of the needed "modifies" clauses. The rest (4%) relied on an Eiffel-specific mechanism, which the analysis does not yet support: redeclaring a function as an attribute in a descendant.

The analysis reported more changed values than specified in the "modifies" clauses. We manually checked that 7 of the inferred clauses indeed reveal unique errors showing discrepancy between specification and implementation. This result is all the more significant that EiffelBase+ is a carefully written library, explicitly designed for formal verification, and has been extensively tested as reported in [16]. (An independent testing effort posterior to the release, using the AutoTest tool for Eiffel, found 5 of the errors, but missed the other two.)

The analysis also detected 7 unnecessary "modifies" specifications: values listed in the specification but not actually changed by the implementation. 4 of these were simply superfluous and could be removed. The remaining 3 were inherited "modifies" specifications; further investigation revealed that they reflected inconsistencies caused by underspecified ancestor contracts.

There were 64 (11%) false positives (clauses inferred but not needed). Of these, 3 were found to reflect actual changes but in unreachable code due to defensive programming in the library. The majority, 46, correspond to the case of a value that the code actually changes, after backing it up, but then restores from the backup. Here the change calculus correctly finds that the value has been changed, twice or more in fact, and other mechanisms are required to find out that the changes cancel each other out. The remaining 15 (2.5% of the total) are the genuine false positives; they are due to the implementation's model of arrays, which does not distinguish between changes to array items and to array size, and which we hope to improve.

The experiments yield the following lessons.

1. Ignoring the temporary problem of functions redeclared into attributes, the change calculus reports 100% of expected "modifies" (frame) properties.
2. It succeeded in pointing out missing "modifies" specifications.
3. It also detected unnecessary "modifies" specifications.

4. The number of false positives is limited, and most of them correspond to values actually changed then restored. Better array handling should entirely limit false positives to this category, plus changes in dead code (which merit attention anyway).
5. If "modifies" specifications rely on model queries (an approach that is not currently dominant, but which we find the most appropriate), the problem remains of mapping attributes to model queries. For the current experiments we performed the mapping manually, but an automatic approach appears possible.

We find these results promising, opening the possibility that automatic alias and change analysis will become a standard component of program verification.

## 6  Future work

Alias analysis can have many applications beyond frame inference. We sketch in subsections 6.1 and 6.2 two other applications, such as deadlock detection and precondition inference. In subsection 6.3 we present a theoretical improvement leading to better performance and precision of alias analysis. This subsection includes a new formal basis for alias calculus and the soundness proof for assignment instruction in this basis. Every subsection reflects work in progress.

### 6.1  Deadlock analysis

The SCOOP concurrency model makes no firm difference between computational mechanisms and resources, all captured by the notion of "processor". For example, in the SCOOP solution of "Dining philosophers", both philosophers and forks are objects residing on their own processors. A processor can access objects handled by another processor by explicitly reserving that object's processor.

The SCOOP reservation mechanism reduces the risk of deadlock by reserving any number of objects atomically, through the syntactical device of argument passing: $r$ ($a$, $b$, …) reserves all the processors of the objects associated with $a$, $b$, …. For example a philosopher will execute *eat* (*left_fork*, *right_fork*). It remains possible, however, to create "Coffman deadlocks" whereby a set of processors reserve each other circularly. The difficulty of detecting them is that processors are known from object references, which may be aliased. Alias analysis may help find possible cycles by considering, in every class, any variable $x$ declared as **separate** (meaning that the object may have a different processor) as aliased to its processor, and looking for cycles in the reservation graph. We are currently implementing this technique.

### 6.2  Precondition inference

In section 4 we saw that the change calculus may lead to inferring "modifies" clauses, conceptually part of a routine's postcondition. This solution is not always appropriate; sometimes we may want to *exclude* the possibility of such change; then the requirement should be in the precondition. The calculus can infer such preconditions automatically. Consider for example the call *a.append* (*b*) where *a* and *b* are both of type

*RESIZABLE_ARRAY*. The routine *append* (*x*: *RESIZABLE_ARRAY*) may have to reallocate the current array to host more elements, and hence to perform

  *area* := *resized_area* (*count* + *x.count*)

which updates the field *area* denoting the place where array elements actually appear. The specification should state that *area* may change; it may do so explicitly through a "modifies" clause, or implicitly through *area*'s appearance in the postcondition as suggested above. But what about *x.area* (*b.area* in the example call)? It could change, for example, if, prior to the call, *a* and *b* are aliased.

It is not necessarily appropriate to add to the "modifies" clause the specification that *x.area* can change; in fact, appending a structure with itself is confusing and a source of bugs. A better solution may be to leave the postcondition of *append* (including the "modifies" property) alone and add to *append* the precondition clause

  *x* /= *Current*

which, in the alias calculus, corresponds to **cut** *x*, *Current*. The alias calculus can help us infer or at least suggest such preconditions automatically.

### 6.3 New mathematical basis: alias diagram model for object-oriented (OO) programming

To improve the rigor of mathematical description and find more effective representation of pointer aliasing properties, we introduce another model based on alias diagrams and show its compatibility with the effect of instructions on object structures, expressed by a semantic model. Depending on the context, the alias relation or alias diagram may be the more convenient view.

The formalization only includes elements relevant to aliasing; in particular, an object contains only reference fields, since value fields such as integers ("expanded" in Eiffel) can be ignored.

The basic sets are **State**, the set of states; **Object**, the finite set of execution objects; **Attr**, the set of tags, corresponding to names of class fields (attributes); **Expr**, the set of expressions; **AD**, the set of alias diagrams. The following conventions are used. For sets $E$ and $F$, sets of functions from $E$ to $F$ are denoted $E \nrightarrow F$ (possibly partial functions), $E \to F$ (total functions) and $E \nrightarrow\!\!\!\!\!\to F$ (functions with a finite domain); the last two are subsets of the first. We generalize lambda notation to partial functions through the convention that if a lambda-expression defines a function $f$, any use of the sub-expression $g(x)$ where $g$ is partial and $x$ is not in its domain implies that the corresponding argument is not in the domain of $f$. Similarly, a predicate is considered true whenever its evaluation involves applying a function outside its domain (i.e. any occurrence of $f(x)$ is implicitly preceded by "$x \in dom\, f \Rightarrow$"). Quantified expressions use a bar |, as in $\forall x: E \mid prop(x)$, to avoid confusion with the dot of OO programming. $\{x: E \mid p(x)\}$ denotes the subset of $E$ made of the elements that satisfy the property $p$. Thanks to variable naming conventions we omit specifying $E$ if it is one of the following sets: **State** (variable names $S, S', S_1 \ldots$); **Object** (names $o, o', o_1, \ldots$).; **Expr**, (names $e, e', e_1, \ldots$); **Attr** (names $t, t', \ldots$); **AD** (names $D, D', \ldots$).

A state is defined as a function returning, for any object and applicable tag, the referenced object (fig. 1):

$$\textbf{State} = \textbf{Object} \to \textbf{Attr} \nrightarrow\!\!\!\!\!\to \textbf{Object}$$

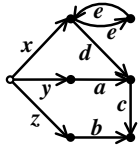

**Fig. 1. An object structure**

An expression (element of **Expr**) is, for this discussion, a path expression: a sequence of zero or more tags separated by dots, as in *a.b.c.d*; *Current* denotes the empty expression. The dot "." can also be used without ambiguity as a concatenation operator on expressions, with the convention that *Current.e* = *e.Current* = *e* for any *e*.

The function *value* defines the semantics of expressions. It has the following signature and recursive definition:

$$value: \textbf{State} \to \textbf{Object} \to \textbf{Expr} \rightharpoonup \textbf{Object}$$
$$value\ (S, o, Current) = o$$
$$value\ (S, o, x.e) = value\ (S, S\ (o, x), e)$$

(For simplicity, $f(a, b)$, for a function $f: A \to B \to C$, means $(f(a))(b)$, and so on.) The presence of the first **Object** in the signature reflects the OO context of this work (see also [13]): an expression is always relative to a "current" object, also known as "this". This object is always considered together with a state (it can be also treated as part of the state but we prefer to pick it out explicitly).

The same property holds for instructions: we understand an instruction as a function *I*: **State** → **Object** ⇀ **State**. In the present work we do not seek a full semantic definition of instructions, but only of their effect on aliasing (to show that the rules of the alias calculus are sound).

It is possible to define that effect using alias relations directly, but a more abstract representation turns out to be more effective: **alias diagrams**. An alias diagram, as illustrated in fig. 2, is a multigraph (labeled directed graph, but with the possibility of more than one edge from a given vertex to another) where: each vertex represents an object (abstracted from execution objects); there is a distinguished vertex called the "root", representing the current object; and every edge is labeled by a tag *x* indicating (in the style of shape analysis [21, 19]) that from any of the objects represented by the source node, the reference *x* can point to one of the objects represented by the target node. **O** *D*, for an alias diagram *D*, is the set of its vertices (objects), **E** *D* the set of its edges and **R** *D* its root. The notion of **path** is the usual one, restricted to finite paths starting at the root. The terminal object of a path *p*, written **T** *p*, is its last vertex (object); its **associated expression X** *p* is the expression $t_1.t_2. \ldots .t_n$ made up of the successive edge tags separated by dots (if more by one). For an empty path, **T** *p* is the root and **X** *p* is *Current*.

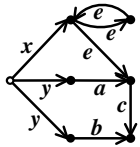

**Fig. 2. An alias diagram**

*Current* represents the empty path from the root to itself. A non-empty path is labeled by an expression, as with *y.a.c* (also expressible as *Current.y.a.c*) in fig. 2.

For an expression $e$, $D\,e \subseteq \mathbf{O}\,D$ denotes the set of terminal objects of paths labeled $e$: $D\,e = \{o \mid \exists\, p \text{ in } D \mid \mathbf{X}\,p = e \wedge \mathbf{T}\,p = o\}$.

An alias diagram represents an alias relation, with the convention that $e$ and $f$ are aliased if and only if one of the following holds: for some node $V$, there are paths labeled $e$ and $f$ both leading to $V$; or $e$ is $e_1.t$ and $f$ is $f_1.t$, where $t$ is a tag and $e_1$ and $e_2$ are expressions (recursively) aliased. As a special case of the first variant, *Current* is aliased to a path $e$ if and only if $e$ leads to the root.

Of most interest are alias diagrams in "canonical form" (closely connected to the canonical form of alias relations seen above), where all vertices are *reachable* and *necessary*. A vertex is **reachable** if there is a path from the root to it (the path may be empty — *Current* — so that the root is always reachable); it is **necessary** if it is the root or has at least two incoming edges or at least one outgoing edge. For the associated alias relation, unreachable and unnecessary vertices are irrelevant; conversely, if a vertex $V$ is reachable and necessary, either $V$ or one of its successors, direct or indirect, has two or more paths leading to it, and hence is relevant for the alias relation. The **canonical form** of a diagram is its maximal canonical sub-diagram.

For any state $S$, if we choose an object $o$ as "current", there is an **associated alias diagram** $D_{S,o}$: the canonical form of the diagram of root $o$ and transitions $S$ (from the definition of **State**, $S$ is indeed a function in **Object** $\to$ **Attr** $\rightharpoonup$ **Object**).

To define the semantics of alias diagrams, we say that a diagram $D$ **holds** in a state $S$ for an object $o$ — written *holds* $(S, o, D)$ — if there is an injective morphism from $D_{S,o}$ to $D$ preserving the root and the transitions. In other words, there must exist a function $i: \mathbf{O}\,D_{S,o} \to \mathbf{O}\,D$ such that

$$(i\,(\mathbf{R}\,D_{S,o}) = \mathbf{R}\,D) \wedge (\forall\, o_1, o_2, t \mid (i\,(o_1) = i\,(o_2) \Rightarrow o_1 = o_2) \wedge$$
$$[o_1, t, o_2] \in \mathbf{E}\,D_{S,o} \Rightarrow [i\,(o_1), t, i\,(o_2)] \in \mathbf{E}\,D)$$

As we have seen, an **alias relation** is a symmetric and irreflexive set of expression pairs $[e, e']$. For an alias diagram $D$ we may obtain the **associated alias relation**:

$$\mathbf{A}\,D = \{[e.e_0, e'.e_0] \mid e \neq e' \wedge (\exists\, p, p': D \mid \mathbf{X}\,p = e, \mathbf{X}\,p' = e', \mathbf{T}\,p = \mathbf{T}\,p')\}$$

(Note that $e$ can be *Current*, in which case $e.e_0$ is just $e_0$, and similarly for $e'$ and $e_0$). $\mathbf{A}\,D$ is the smallest set containing $\{[e, e'] \mid e \neq e', \exists\, p, p' \text{ in } D \mid \mathbf{X}\,p = e, \mathbf{X}\,p' = e', \mathbf{T}\,p = \mathbf{T}\,p'\}$ and satisfying the dot completeness condition: for any $[e, e']: \mathbf{A}\,D$ and $x:$ **Attr**, then $[e.x, e'.x] \in \mathbf{A}\,D$. For soundness, the following property should hold:

$\forall\, S \mid \forall\, o \mid \forall\, D \mid$ *holds* $(S, o, D) \Rightarrow$
$\qquad (\forall\, e, e' \mid e \neq e' \wedge$ *value* $(S, o, e) =$ *value* $(S, o, e') \Rightarrow [e, e'] \in \mathbf{A}\,D)$ (1)

This definition reflects the conservative (over-approximation) nature of the alias calculus: it states that if different expressions $e$ and $e'$, defined relative to an object $o$ and a state $S$, have the same value (point to the same object), then the pair $[e, e']$ must be in the alias relation for the state to which the object belongs. There is no reverse implication, which would correspond to a "must-alias" analysis.

The alias relations associated with an alias diagram and its canonical form are the same. Also, *holds* $(S, o, D_{S,o})$ is always satisfied.

The alias calculus is a set of rules that for any instruction *I* and alias diagram *D* yield another alias diagram *D » I*. In this framework, soundness is defined as:

$$\forall S \mid \forall o \mid \forall D \mid holds\,(S, o, D) \Rightarrow holds\,(I\,(S, o), o, D » I) \qquad (2)$$

The soundness proof must establish (1) and show that for every kind of instruction *I* the corresponding rule in the alias calculus, as given in section 3, satisfies (2).

Now we're ready to discuss a soundness proof for the assignment rule. This rule was chosen because it is at the core of the calculus and because, as stated in section 3, previous versions [11, 12] were slightly erroneous.

An assignment *t := e*, where *t* (the target) is a tag and *e* (the source) an expression, denotes the following function:

$$\lambda S \mid \lambda o \mid S - \{[o, t, value\,(S, o, t)]\} \cup \{[o, t, value\,(S, o, e)]\}$$

where "–" is set difference. In words: to obtain the new state, first remove the value associated with the target tag *t* for the target object *o*, then add, for that tag, the value of the source expression *e* for *o* in the original state *S*.

Given an expression $e = t_1.t_2...t_n$, the alias rule of section 3 can be expressed, in terms of alias diagrams rather than alias relations, as

$$\mathbf{O}\,(D » t := e) = \mathbf{O}\,D \cup \{v_1, ..., v_n\}$$
$$\mathbf{E}\,(D » t := e) = \mathbf{E}\,D \cup \{[o, t_{k+1}, v_{k+1}] \mid o \in D\,t_1.t_2...t_k, 0 \leq k < n\}$$
$$\cup \{[v_k, t_{k+1}, v_{k+1}] \mid 1 \leq k < n\}$$
$$- \{[\mathbf{R}\,D, t, o] \mid o: \mathbf{O}\,D\} - \{[\mathbf{R}\,D, t, v_1]\}$$
$$\cup \{[\mathbf{R}\,D, t, o] \mid o \in D\,e\} \cup \{[\mathbf{R}\,D, t, v_n]\}$$
$$\mathbf{R}\,(D » t := e) = \mathbf{R}\,D$$

In words: for a diagram *D* add a new path corresponding to expression *e*, and for every vertex corresponding to any prefix of *e* add an edge from it to the vertex of this new path corresponding to the next prefix of *e*. Then remove all the edges labeled *t* from the root and add edges labeled *t* from the root to all vertices corresponding to expression *e* in *D* and to the last vertex of the new path.

**Theorem** to be proved: (2) holds if *I* is an assignment instruction "*t := e*" and *D » I* is defined by the assignment axiom above.

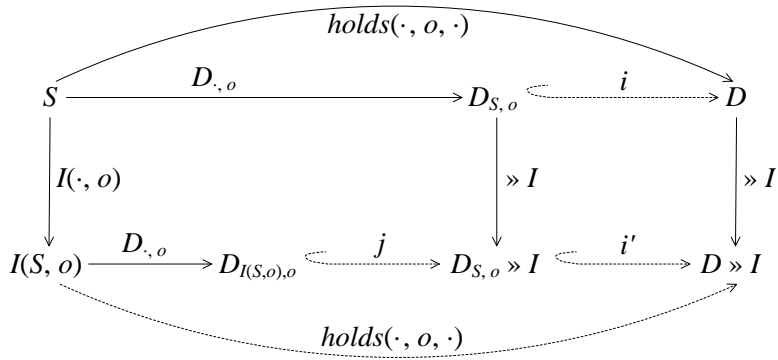

**Fig. 3. Soundness of the assignment rule**

The proof now follows. Let *S* be a state, *o* an object, *D* an alias diagram, *I* the assignment "*t := e*", and assume that *holds (S, o, D)* is satisfied. For the alias diagram $D_{S,o}$ associated with *S*, *holds* (*S*, *o*, $D_{S,o}$) is trivially satisfied. We also know from the assumption and the definition of *holds* that there is an injective function *i*: **O** $D_{S,o}$ → **O** *D* preserving the root and transitions. The existence of *i* means that $D_{S,o}$ can be embedded into *D* via *i*. The operation "» *I*" is monotone (for detailed discussion of alias monotonicity see [12]). Monotonicity implies that there is a similar injection *i'* between the object sets of the diagrams $D_{S,o}$ » *I* and *D* » *I* (fig. 3).

Next, consider two alias diagrams $D_{I(S,o),o}$ and $D_{S,o}$ » *I*. From the construction of the associated alias diagram and the definition of the assignment rule it follows that there is an injection *j*: **O** $D_{I(S,o),o}$ → **O** ($D_{S,o}$ » *I*). Since *holds* (*I* (*S*, *o*), *o*, $D_{I(S,o),o}$) is trivially satisfied, the composition *i'* ∘ *j* satisfies the *holds* definition: the existence of such injection implies *holds* (*I* (*S*, *o*), *o*, *D* » *I*) (fig. 3).

We checked this proof of the assignment rule mechanically using the Coq proof assistant. The proof is available at sel.ifmo.ru/results/alias/semantics/.

## 7 Related work

There is a considerable literature on alias analysis, in particular for compiler optimization. We only consider work that is directly comparable to the present approach.

### 7.1 Alias analysis rules

There are different approaches to compute alias information for programs. All of them, including classic iteration-based variants converging to a fixed point and equation-based techniques as in [14], define a set of rules that help compute alias information. The rules are associated with program elements, expressions and instructions (statements), and specify how they affect the model elements used to compute alias information. In C-like languages this usually includes [5, 8, 14]:

*Address-of* / *Alloc* (*y* = &*x*)              *Copy* (*x* = *y*)
*Load* (*y* = \**x*)                           *Store* (\**x* = *y*)

Here we only mention the differences for the assignment instructions, but depending on the level of the language there could be some more instructions and associated rules. For example, [18] uses an intermediate language, RTL, to perform the analysis.

Many of the earlier approaches address C or languages of that level; the present work has been applied to a full-fledged object-oriented language. In an OO context some of the instructions may become unnecessary. In particular, there is no notion of plain pointers. They are replaced by class fields [23]:

*New* (*x* := new *O*)                 *Assign* (*x* := *y*)
*Load* (*y* := *x.f*)                      *Store* (*x.f* := *y*)

The rules can be simplified even further when an OO language, such as Eiffel, in line with the information hiding principle, disallows remote modification of an object field (*x.u* := *a* must be written *x.set_u* (*a*) using a setter *set_u*); then there is no need for *Store*. The formalism and implementation of the present work rely on that assumption but can be generalized to languages a language accepting direct.

Many earlier approaches are flow-insensitive: in *a* := *b*; *a* := *c* they will find that *a* can be aliased to both *b* and *c*. Such imprecision is unacceptable for the applications examined in the present work, such as change analysis and frame inference. An example of flow-sensitive analysis is [5], but it too introduces imprecision, in particular in its handling of assignment. As compared to such work the high precision of our approach is obtained at the expense of performance, although we hope to improve it.

The analyses of which we are aware compute the alias information using only instructions that may change the object state or a variable value. It is also useful to introduce constructs that do not change any state, but do change the alias information; they include, as described in sections 3 and 6.2, instructions asserting equality such as **cut** asserting $x \mathrel{/{=}} y$. (Our work also uses **bind**, which asserts $x = y$.) We do not know of other alias work using such instructions. They make it possible to take advantage of Hoare-style assertions for alias analysis, rather than simply ignoring them, and may provide a way to combine may-alias and must-alias analysis as suggested in [3].

As in some other inter-procedural analyses [3, 4], the information computed for every routine is recorded for later use to avoid unnecessary recomputation.

### 7.2   Soundness proof

This paper presented a proof of soundness in section 6.3. A soundness proof for alias analysis appears in [18], using Coq as in the present work. The proof in [18], however, applies to C, through an intermediate language. The proof presented here, and the underlying theory (alias calculus), apply directly to the programming language. In addition, that programming language is not C but an OO language.

### 7.3   Frame condition inference

Automated support for code verification is a well-known problem that has been tackled in the past two decades with increasing success. The approaches range from static analysis as in [20] to dynamic contract inference as in [15]. Instead of the contracts in general here we focus on frame conditions. This is similar to the work described in [17], but it is done in the context of a safe object-oriented language.

According to [16], the completeness of the contracts is an important condition for realistic program verification. The contracts should include not only pre- and postconditions but also "modifies" clauses that list all the data affected by the particular method of a class. It turns out [17] that 90% of such information can be obtained automatically. Unlike other methods, our analysis uses available pre- and postconditions to avoid inferring unnecessary "modifies" clauses.

The frame conditions could also be proved with must-alias analysis, or even by applying may-alias and must-alias analyses together as described in [3]. For every attribute $x$ of a class the following property could be checked at routine exit: $x = $ **old** $x$, where **old** $x$ stands for the value of $x$ on routine entry. Indeed, must-alias analysis would tell whether this expression is always true. But then the problem is to build an exhaustive list of all the (possibly nested) attributes of all the reachable objects and to apply must-alias analysis to this list that does not seem practical.

# 8    Conclusion

The alias calculus and change calculus, as described here, are implemented as part of the EVE development environment; the reader can try them out by downloading EVE at the URL given in [1]. A number of challenges remain open:

- Building alias calculus rules for composite constructions in the model of alias diagrams. The rules must be sound and allow efficient implementation.
- Close integration with other verification tools, in particular (in EVE) the Boogie-based AutoProof proof system (taking advantage of the interplay, discussed in section 3, between the automatic alias calculus and annotation-based Hoare-style proofs), and the AutoTest automatic testing mechanism.
- New applications, including deadlock detection, as sketched in section 6.1.
- Better integration of modularity concerns; although the calculus supports modularity, the current implementation has not focused on this aspect.
- Performance improvement; 7 minutes per class is acceptable for an initial version, especially if the tools run in the background, but turning change and alias calculus into routine tools of the environment, with immediate feedback, requires a significant performance improvement.
- Human engineering, in particular the development of suitable mechanisms to display the results of alias and change analysis in a form directly meaningful for programmers, and as a tool for suggesting preconditions and other missing contracts.

Among the main benefits of the approach as developed so far, we find the following: it is entirely automatic (with the provision of cut and bind as escape mechanisms); it is of much higher precision than many of the existing approaches (the only sources of imprecision being the neglect of conditionals and the approximation of infinite diagrams by finite but large ones); it is based on a simple and (we hope) convincing calculus; its soundness has been partly established; it applies to a full-fledged, practical, modern OO language; and it is implemented as part of a modern IDE. We believe the approach provides a significant practical advance towards the automatic computation of frame properties and other fundamental program properties resulting from the unpleasant but inevitable presence of aliasing in modern programming frameworks.